\documentclass[graybox]{svmult}

\usepackage{mathptmx}       
\usepackage{helvet}         
\usepackage{courier}        
\usepackage{type1cm}        
\usepackage{makeidx}         
\usepackage{graphicx}        
\usepackage{multicol}        
\usepackage[bottom]{footmisc}
\usepackage{amsmath}

\makeindex             

\begin{document}

\title*{Quantum Simulation of Spin Chains Coupled to Bosonic Modes with Superconducting Circuits}
\titlerunning{Spin Chains Coupled to Bosonic Modes with Superconducting Circuits}
\author{U. Las Heras$^1$, L. Garc\'ia-\'Alvarez$^1$, A. Mezzacapo$^{1,\dag}$, E. Solano$^{1,2}$ and L. Lamata$^1$}
\institute{$^1$Department of Physical Chemistry, University of the Basque Country UPV/EHU, Apartado 644, E-48080 Bilbao, Spain\\
$^2$IKERBASQUE, Basque Foundation for Science, Maria Diaz de Haro 3, 48013 Bilbao, Spain\\\\
$^\dagger$Now at: IBM T. J. Watson Research Center, Yorktown Heights, NY 10598, USA}

\authorrunning{U. Las Heras et al.}


%
%
\maketitle

\abstract*{Abstract of the simulation of the spin models in CQED.}

\abstract{We propose the implementation of a digital quantum simulation of spin chains coupled to bosonic field modes in superconducting circuits. Gates with high fidelities allows one to simulate a variety of Ising magnetic pairing interactions with transverse field, Tavis-Cummings interaction between spins and a bosonic mode, and a spin model with three-body terms. We analyze the feasibility of the implementation in realistic circuit quantum electrodynamics setups, where the interactions are either realized via capacitive couplings or mediated by microwave resonators.}

\section{Introduction}
\label{sec:1}
A two-level system coupled with a single radiation mode is modeled by the ubiquitous and paradigmatic quantum Rabi model~\cite{Rabi36} that describes the most fundamental interaction between quantum light and quantum matter. There have been many efforts, both in theory and experiments, to capture the features of this model in different quantum technologies~\cite{Lewenstein12,Nori14}. These analysis have an impact on understanding about different quantum phenomena~\cite{Casanova10,Koch10,Niemczyk10,Oudenaarden96,Rotondo15,You10}.

The concept of a quantum simulator can be attributed to Feynman~\cite{Feynman82}, and it refers to a controllable quantum platform that mimics the behaviour of another quantum system. Analog quantum simulators have been proposed and implemented in several quantum technologies, such as trapped ions~\cite{Casanova12,Mezzacapo12}, ultracold atoms~\cite{Bloch12}, or superconducting circuits~\cite{Ballester12,Pedernales13,Viehmann13,Mostame15}. Digital methods based on discrete-time gate sequences~\cite{Lloyd96} in order to simulate dynamics of quantum systems have been proposed and realized in trapped ions~\cite{Lanyon11}, photonic systems~\cite{LanyonQChem}, spin-photon hybrid systems~\cite{Chiesa15} and superconducting circuits~\cite{Salathe15,Barends15,Heras14,Mezzacapo14,Heras15,Geller12}.

Here, we analyze the quantum simulation of arbitrary and generic models, where spin chains alone or coupled to bosonic modes are emulated in superconducting circuits~\cite{Devoret13}. We use digital techniques in order to imitate systems whose dynamics in principle may differ from the ones of the experimental setups. Finally, we study the feasibility and efficiency of the implementation of three generic models in a realistic circuit quantum electrodynamics setup.

\section{Digital Quantum Simulations}
\label{sec:2}
The goal of simulating diverse and generic models involving spin interactions and bosonic modes leads us to consider digital techniques, due to their suitability and flexibility for mimicking different dynamical structures. Hamiltonian dynamics can be approximated by the digital decomposition of the exact unitary evolution into discrete stepwise unitary operations, implemented by using quantum gates in an efficient way~\cite{Lloyd96,Suzuki90}. Digital methods are based on the Trotter formula, which allows us to expand the evolution operator of Hamiltonians that are written as a sum of terms, $H = \sum^N_{j=1} H_j$, into a product of evolution operators for the interactions given by the summands of the Hamiltonian, $H_j$. The Trotter expansion can be written as
\begin{equation}
e^{-iHt} = \left(e^{-iH_1 t/s} \cdots e^{-iH_N t/s}\right)^s + \sum_{i<j}\frac{\left[H_i,H_j\right] t^2}{2s} + \sum_{k=3}^\infty E(k),\label{EqTrotter}
\end{equation}
where the total time of the simulated dynamics is divided into $s$ intervals in which the evolution associated to each summand of the complete Hamiltonian are applied. The error scales with $t^2/s$ for short times, as can be observed in the second term, and the upper bound for higher order error contributions is $s\|Ht/s\|^k_{\rm sup}/k! \ge \|E(k)\|_{\rm sup}$.

Our goal is to propose a systematic procedure using digital methods for simulating efficiently different models, namely spin-spin interaction and spins coupled to bosonic modes. First, we employ gates that commute with each other and do not produce digital error. For those that do not commute, we apply several Trotter steps because the more Trotter steps one applies, the smaller the digital error produced is. In realistic experiments, one has to take into account decoherence times and gate errors. Therefore, we have to regulate the number of steps in order to be able to perform the simulation before decoherence effects take place, and in order to reduce the accumulated gate error. Consequently, once the digital error is small enough applying a certain number of Trotter steps, the error coming from the experimental setup always must be smaller than the digital one.

\section{Quantum simulation of spin chains coupled to bosonic modes with superconducting circuits}
\label{sec:3}
In this section, we present a method to implement the dynamics of several spin models, coupled with bosonic modes, in circuit quantum electrodynamics setups. For this purpose, we take under consideration two different architectures of superconducting circuits. We show how to use linear arrays of superconducting qubits with capacitive coupling between nearest neighbors~\cite{Barends14} to simulate the Ising model with transverse field. Then we simulate the behavior of a spin-chain coupled to a bosonic mode via a Tavis-Cummings interaction~\cite{Tavis68}. Moreover, we show how to implement many-body spin dynamics with highly nonlinear terms in superconducting qubits coupled to transmission line resonators acting as a quantum bus~\cite{Wallraff04}.

\begin{figure}
\sidecaption[t]
\includegraphics[scale=.45]{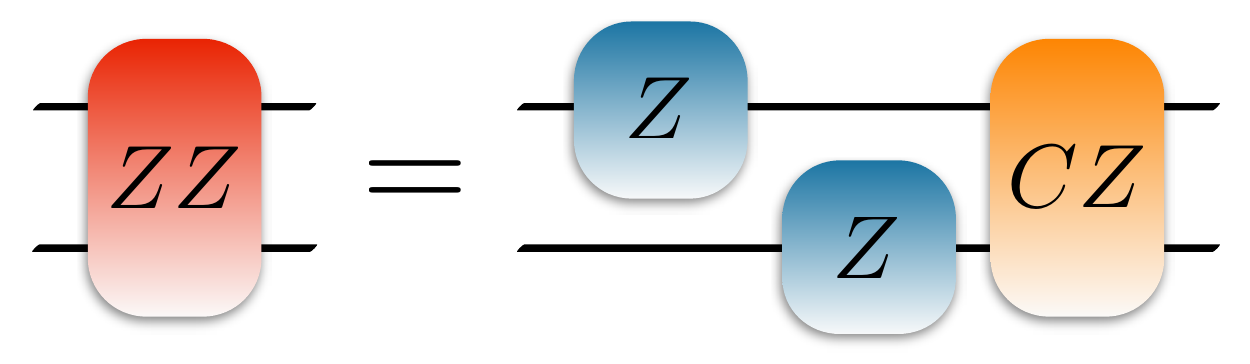}
%
%
\caption{Protocol for decomposing a $ZZ$ interaction between two qubits using single-qubit rotations, $Z$, and a c-phase gate, $CZ$.}
\label{FigZZProtocol}       
\end{figure}

In the following, we propose digital quantum simulations based on quantum gates implemented in superconducting architectures. Capacitive coupling setups allow one to implement $ZZ$ gates, $\exp(-i\theta\sigma^z_j\sigma^z_k)$, for nearest-neighbor superconducting qubits by the sequence of two single qubit rotations along the $z$ axis, $Z(\phi)$, and a c-phase gate, $CZ(\phi)$, as shown in Fig.~\ref{FigZZProtocol}, where 
\begin{eqnarray}
Z(\phi)=\left(\begin{array}{cc} 1 & 0 \\0 & e^{i\phi}\end{array}\right)\!\!, \,\,CZ(\phi)=\left(\begin{array}{cccc} 1 & 0 & 0& 0\\0 & 1 & 0 & 0\\ 0 & 0 & 1 & 0\\ 0 & 0 & 0 & e^{-i2\phi}\end{array}\right)\!\!.\label{EqCZ}
\end{eqnarray}
The current achievable fidelities in superconducting circuits are of~\cite{Barends14} $99.9\%$ and $99.4\%$ for the single and two-qubit (CZ) gates, respectively. They enable circuit QED setups to be great candidates for digital quantum simulators where the stroboscopic application of gates is necessary. Notice that $ZZ_{12}(\theta)=(Z_1(\phi)\otimes Z_2(\phi))CZ_{12}(\phi)$ for $\theta=\phi/2$.

The use of quantum buses allows for the implementation of multi-qubit gates and spin-boson interactions, coupling the electromagnetic field in the resonator with superconducting circuits~\cite{Mezzacapo14,Garcia-Alvarez14,Mei13,Seo15}.

\subsection{Ising model with transverse field via capacitive nearest-neighbour gates}
\label{sec:3.1}

One of the most studied spin models in condensed matter is the Ising model with a transverse field, which is used for describing the behavior of interacting nearest-neighbor dipoles in the presence of a transverse magnetic field. The Hamiltonian of $N$ spins can be written as
\begin{equation}\label{ITF}
H_{ITF}=J\sum_{\langle jk \rangle}\sigma^z_j\sigma^z_k+B\sum_j\sigma^x_j,
\end{equation}
where $\sigma_j^\alpha$ is the Pauli operator acting over the $j$-th spin with $j=1,...,N$, in the direction $\alpha=x,y,z$. $J$ stands for the coupling between nearest-neighbor spins and $B$ is the coupling between a spin and the transverse field. Depending on the sign of $J$ the model is ferromagnetic $(J<0)$ or antiferromagnetic $(J>0)$. In order to reproduce this interaction in superconducting circuits, we make use of a high-fidelity set of gates, as introduced in Eq.~(\ref{EqCZ}): single-qubit rotations along the $x$ direction, $X_j(\phi)=\exp(-i\phi\sigma^x_j)$, and two-qubit $ZZ$ gates, $ZZ_{jk}(\theta)=\exp(-i\theta\sigma^z_j\sigma^z_k)$.

\begin{figure}
\sidecaption[t]
\includegraphics[scale=.45]{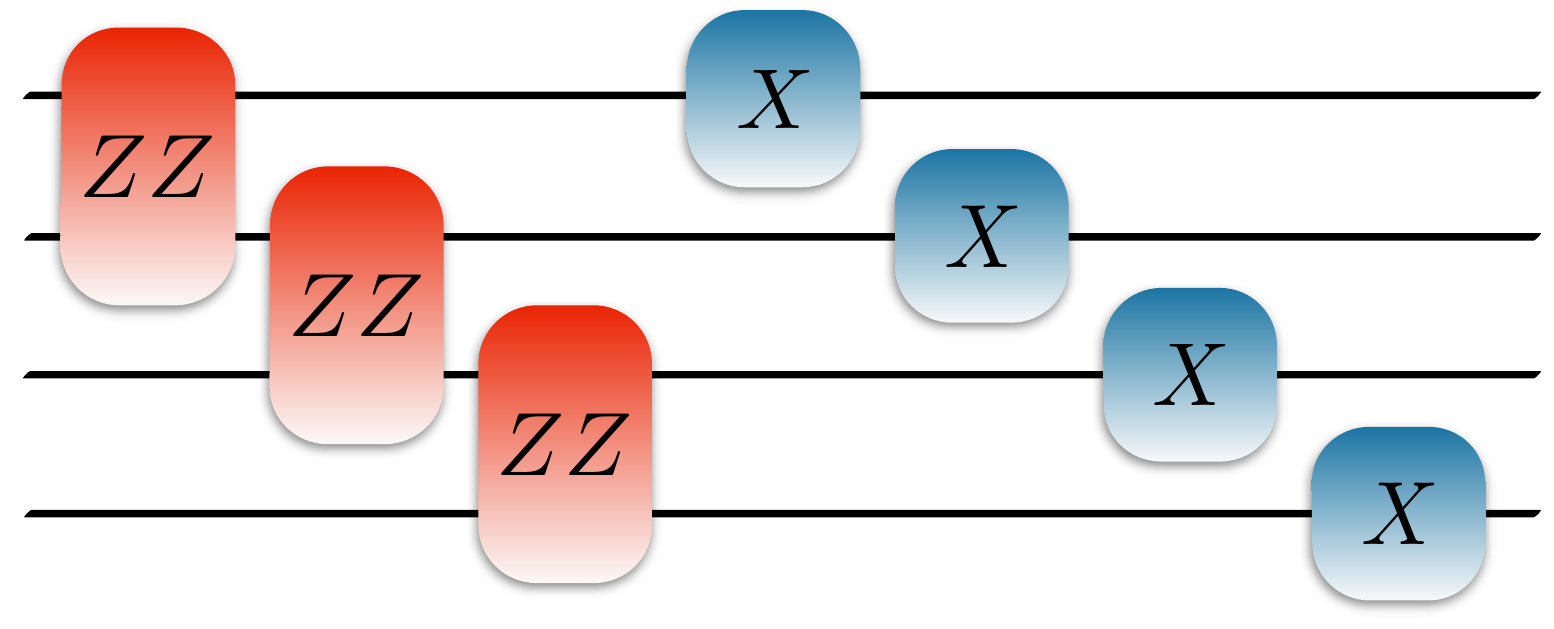}
%
%
\caption{Protocol for digital quantum simulation of the Ising model with transverse magnetic field in terms of $ZZ$ two-qubit gates and single qubit rotations along $x$ axis.}
\label{FigIsingProtocol}       
\end{figure}

As shown in Section~\ref{sec:2}, it is possible to decompose a complex interaction into discrete series of gates through a Trotter expansion. In order to implement the spin-spin interaction, we need to execute $(N-1)$ two-qubit gates. In this case, there is no digital error because all the gates in this decomposition commute,
\begin{equation}
\label{twoqubit}
\exp(-i\ \theta\sum_{\langle jk \rangle}\sigma^z_j\sigma^z_k)=e^{-i\theta\sigma^z_1\sigma^z_2}\ e^{-i\theta\sigma^z_2\sigma^z_3}\ \cdots\ e^{-i\theta\sigma^z_{N-1}\sigma^z_N},
\end{equation}
with $\theta=Jt$, $t$ being the simulation time of the experiment.The coupling among the spins and the transverse field can be simulated in a similar way using $N$ single qubit rotations,
\begin{equation}
\label{singlequbit}
\exp(-i\ \phi\sum_j\sigma^x_j)=e^{-i\phi\sigma^x_1}\ e^{-i\phi\sigma^x_2}\ \cdots\ e^{-i\phi\sigma^x_N},
\end{equation}
with $\phi=Bt$. Given that the two interactions in Eqs.~(\ref{twoqubit}) and (\ref{singlequbit}) do not commute, one has to implement them in sequential short-time Trotter steps to minimize the digital error. In Fig.~\ref{FigIsingProtocol}, we show a scheme of the protocol for the quantum simulation of the transverse field Ising model for four spins. The recent achievement of high-fidelity single and two-qubit (CZ) gates with superconducting circuits will allow one to perform many Trotter steps for several qubits, using hundreds of gates.

\begin{figure}
\sidecaption[t]
\includegraphics[scale=.68]{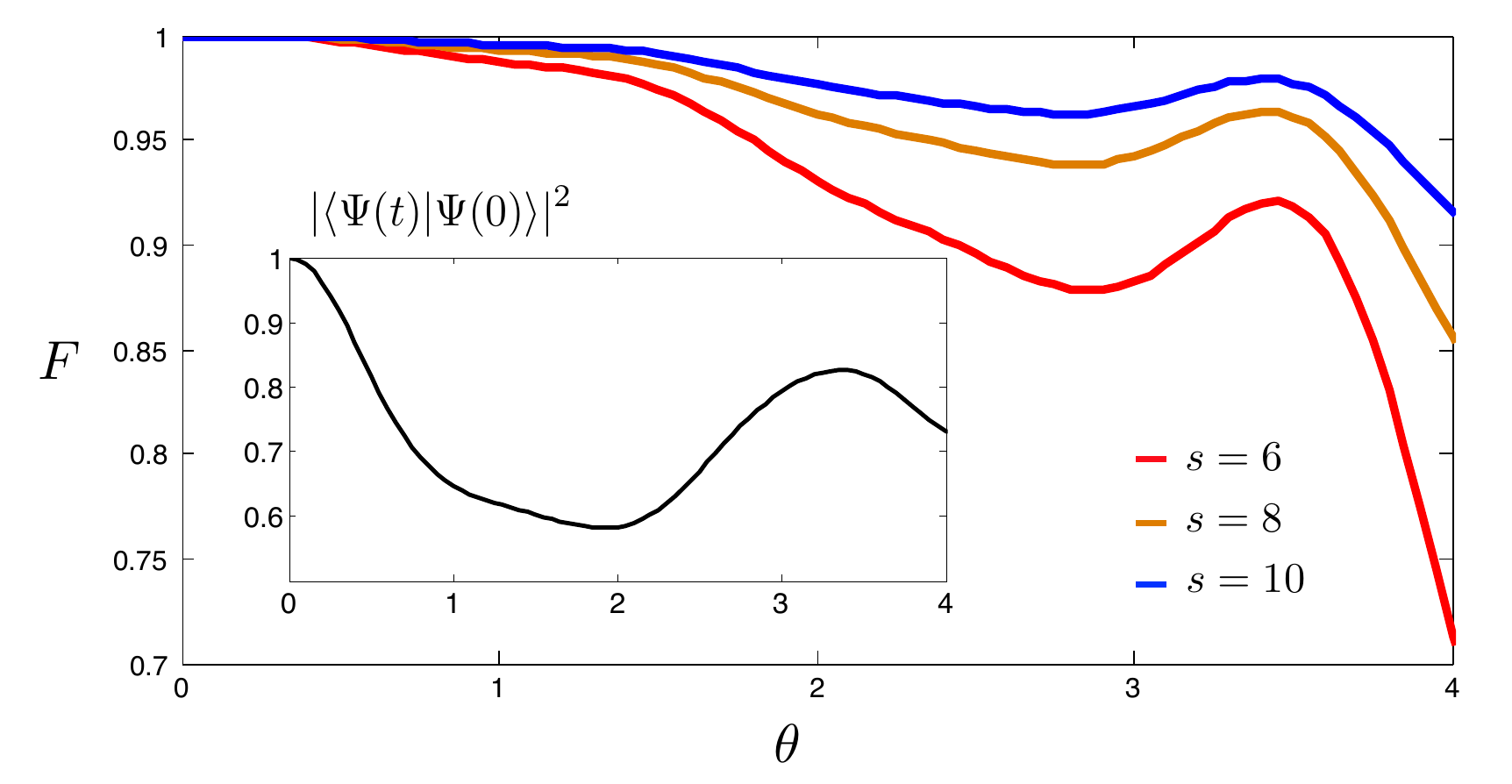}
%
%
\caption{Digital simulation of the ferromagnetic Ising model with a transverse field for four spins in superconducting circuits, up to a phase of $\theta=4$, with $J/B=2$. The plot shows the fidelity of the digitally evolved state versus the ideally evolved one for different number of Trotter steps, $s=6,8,10$. The inset shows the overlap between the ideally evolved state with the initial state, that is, all qubits in $|0\rangle_z$.}
\label{FigIsingPlot}       
\end{figure}

In order to reduce the digital error, it is necessary to increase the number of Trotter steps. In Fig.~\ref{FigIsingPlot}, we plot a numerical simulation of the Ising model with transverse field for different digital steps. The simulated dynamics with digital decomposition is more accurate when compared with the exact dynamics when the number of Trotter steps is increased. 

\subsection{Spin chain coupled to a bosonic field mode via Tavis-Cummings interaction}
\label{sec:3.2}

We now analyze a model consisting of a spin-chain with nearest-neighbour couplings interacting with a bosonic mode. In this sense, both free energies of the bosonic mode and spins are taken into account, as well as spin-spin and spin-boson interactions. The spin-spin evolution is modelled with the Ising dynamics, while the Tavis-Cummings model describes the interactions between spins and bosons. The resulting Hamiltonian is
\begin{equation}
H_{ITC}=\omega\ a^\dagger a+\sum_{j}\frac{\Omega}{2}\sigma^z_j-J\sum_{\langle jk \rangle}\sigma^z_j\sigma^z_k+g\sum_j(a\sigma^+_j+a^\dagger\sigma^-_j).
\end{equation}
Following the notation presented above, $\sigma^z_j$ is the Pauli operator along $z$ direction, $\sigma_j^+(\sigma_j^-)$ is the creation(annihilation) spin excitation operator acting on the $i$-th spin and $a(a^\dagger)$ is the annihilation(creation) operator of the bosonic mode. $\omega$ and $\Omega$ are the free energies of each boson and spin, respectively. Moreover, $J$ is the coupling constant between nearest spins and $g$ stands for the coupling among spins and bosonic field.

The implementation in circuit QED requires the simulation not only of the spin dynamics, as in the previous example, but also of the bosonic mode. To achieve this, the superconducting qubits play the role of spins while the photons in a transmission line resonator emulate the bosonic excitations in the simulation. In order to perform the interactions of the model, it is necessary to couple the resonator to all the superconducting qubits. The Tavis-Cummings interaction appears straightforwardly in circuit QED setups once the rotating wave approximation is performed,
\begin{equation}
H_{1}=\omega_1\ a^\dagger a+\sum_{j}\frac{\Omega_1}{2}\sigma^z_j+g\sum_j(a\sigma^+_j+a^\dagger\sigma^-_j),
\end{equation}
being $\omega_1$ the frequency of the photons in the resonator, $\Omega_1$ the frequency associated with the excitation energy of the superconducting qubits, and $g$ the qubit-resonator coupling constant. The spin-spin interaction for qubits $j$ and $k$ is achieved by means of the $ZZ$ gate presented in Eq.~(\ref{EqCZ}). Detuning to high frequencies the qubit-resonator interaction we are able to reproduce the model
\begin{equation}
\label{Htilde}
H(j,k)=\omega'\ a^\dagger a+\sum_{j}\frac{\Omega'}{2}\sigma^z_j-J\sigma^z_j\sigma^z_k.
\end{equation}
Since $[H(j,k),H(j',k')]=0\ \forall\ j,j',k,k'$, we can define and implement sequentially the interaction
\begin{equation}
H_2=\sum_{\langle jk \rangle} {H}(j,k)=\omega_2\ a^\dagger a+\sum_{j}\frac{\Omega_2}{2}\sigma^z_j-J\sum_{\langle jk \rangle}\sigma^z_j\sigma^z_k,
\end{equation}
where $\omega_2=(N-1)\omega'$, $\Omega_2=(N-1)\Omega'$ and $N$ the number of simulated spins, and it fulfills the condition $\exp(-itH_2)=\prod_{\langle jk\rangle} \exp(-it H(j,k))$, being $t$ the execution time.

\begin{figure}
\sidecaption[t]
\includegraphics[scale=.68]{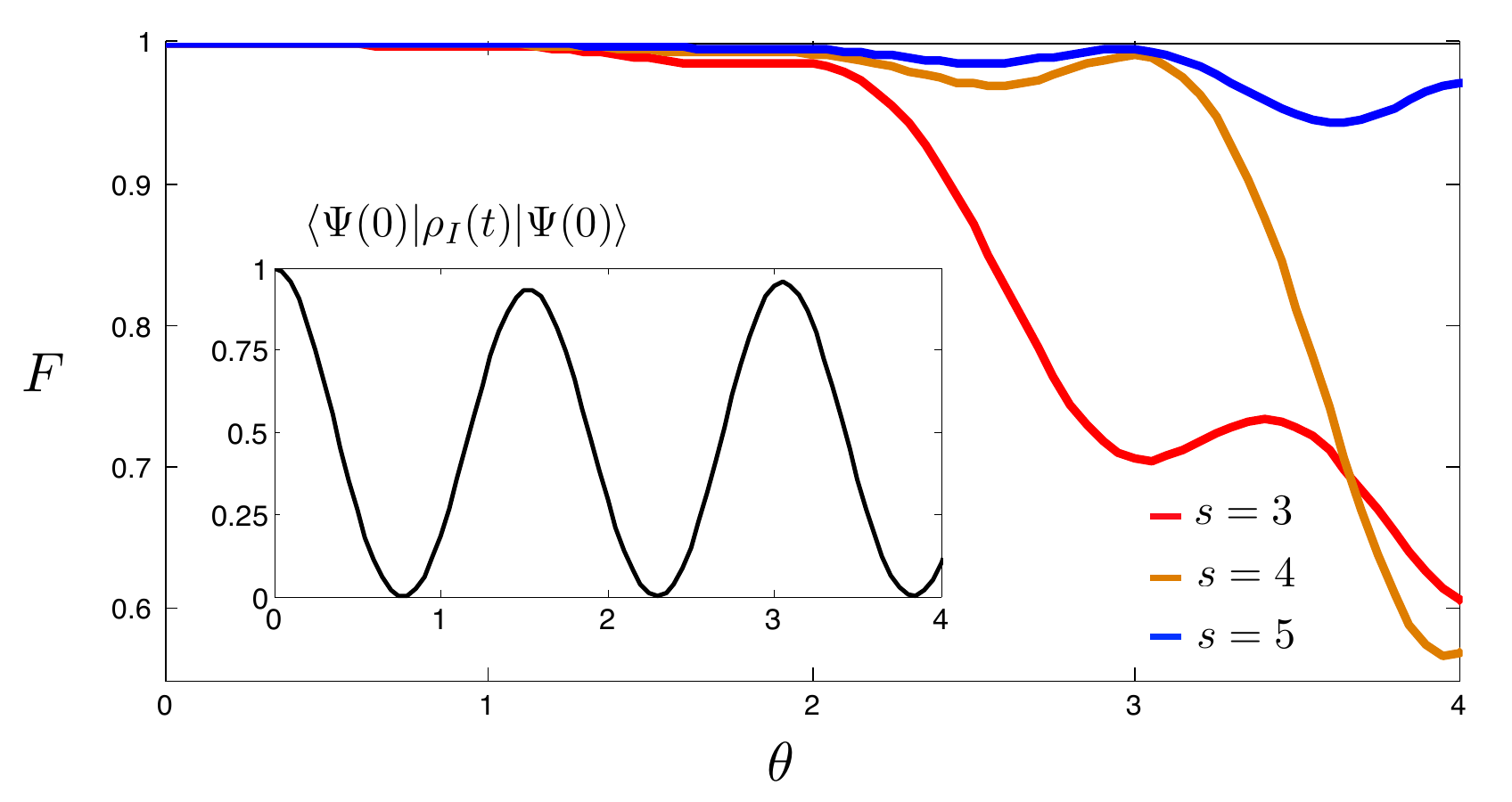}
%
%
\caption{Fidelity $F$ of the simulation of a four-spin chain coupled to a bosonic mode with circuit QED for different Trotter steps, $s=3,4,5$. The upper curves correspond to larger number of Trotter steps. Here, the parameters of Hamiltonians $H_1$ and $H_2$ are $\omega_1=2\pi\times200$~MHz, $\Omega_1=2\pi\times180$~MHz, $g=2\pi\times80$~MHz, $\omega_2=2\pi\times600$~MHz, $\Omega_2=2\pi\times18$~MHz, $J=2\pi\times200$~MHz and the decay rate of the resonator is given by $\kappa=2\pi\times10$~kHz. $F$ is defined as the overlap between the ideally evolved density matrix and the digitally evolved one, $F(t)={\rm Tr} (\rho_I(t)\rho_T(t))$. The inset shows the overlap between the ideally evolved density matrix and the state of the system at $t=0$, $1/\sqrt{2}(a^\dagger+(a^{\dagger})^2/\sqrt{2})|0\rangle_p\otimes|1_10_20_30_4\rangle_z$, i.e., the same probability for having 1 and 2 photons in the resonator and all the superconducting qubits in the ground state of $\sigma^z_i$ except the first, which is excited.}
\label{FigITCPlot}       
\end{figure}

\begin{figure}
\sidecaption[t]
\includegraphics[scale=.5]{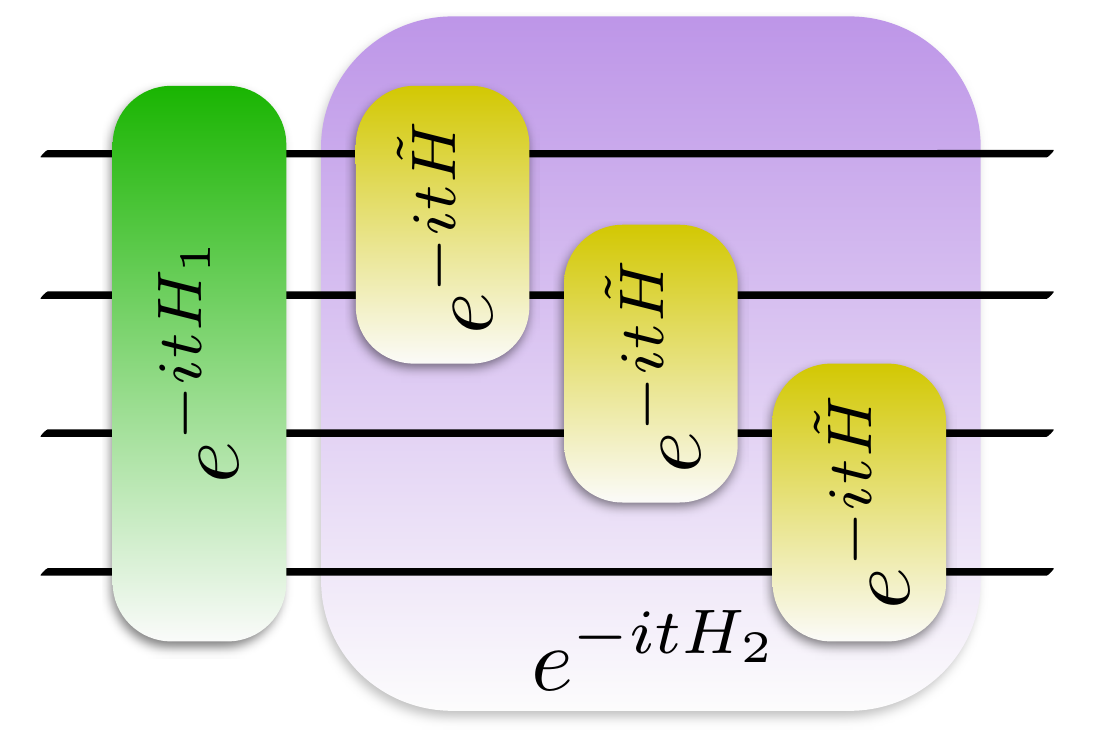}
%
%
\caption{Protocol for the digital quantum simulation of a spin-chain coupled to a bosonic mode with superconducting circuits, in terms of unitary evolutions of Hamiltonians $H_1$, $H_2$ and $\tilde{H}$, being the interaction defined in Eq.~(\ref{Htilde}).}
\label{FigITCProtocol}       
\end{figure}

Summing the interactions $H_1$ and $H_2$ we recover the model we wanted to reproduce, $H_{ITC}$. Nevertheless, $[H_1,H_2]\neq0$, so we need to employ the Trotter method in order to make the digital error decrease, as shown in Fig.~\ref{FigITCPlot}. Moreover, for considering the resonator photonic leakage, we have calculated the evolution of the system making use of the master equation,
\begin{equation}
\dot{\rho}=-i[H_{t},\rho]+\kappa L(a)\rho,\label{master}
\end{equation}
where $L(a)\rho=(2a\rho a^{\dagger}-a^{\dagger}a\rho-\rho a^{\dagger}a)/2$ is the Lindblad superoperator acting on $a$, $\kappa$ is the decay rate of the resonator, and $H_t=\{H_1,H_2\}$ is the Hamiltonian that governs the evolution. Notice that we have considered a coherence time much longer for the qubits than for the resonator. In Fig.~\ref{FigITCProtocol}, we plot the steps for implementing the protocol for four spins interacting with a bosonic mode.

\subsection{Collective spin coupling mediated by resonators}
\label{sec:3.3}
\begin{figure}
\sidecaption[t]
\includegraphics[scale=.5]{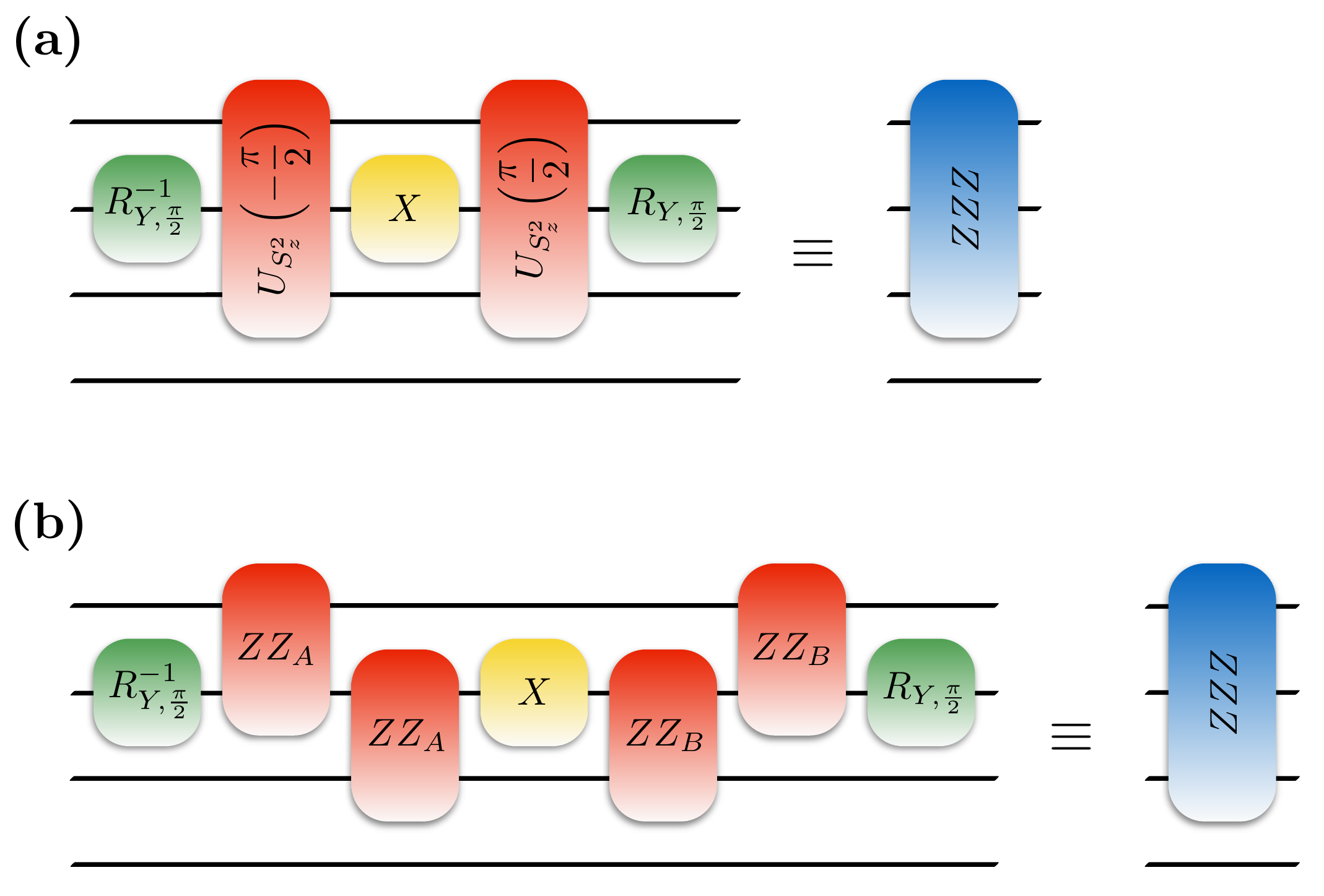}
%
%
\caption{ {\textbf{a}} Protocol for performing one of the three-qubit interactions appearing in Eq.~(\ref{CSC}) with collective gates. Here $ZZZ_{123}(\alpha) = \exp(-i \alpha \sigma^z_1\sigma^z_2\sigma^z_3)$, $R_{Y,\theta} = \exp(-i \theta \sigma^y /2)$ is the rotation along the $Y$-axis of a qubit, $X = \exp(i \alpha \sigma^x)$, and $U_{S_z^2}(\theta)=\exp(-i\theta/2\sum_{i<j}\sigma^z_i\sigma^z_j)$. {\textbf{b}} The same interaction $ZZZ$ can be realized with two-qubit gates where $ZZ_A = \exp(i \pi \sigma^z \otimes \sigma^z /4)$, and $ZZ_B = \exp(-i \pi \sigma^z \otimes \sigma^z /4)$.}
\label{MSProtocol}       
\end{figure}

In this subsection, we extend the Ising model presented in Eq.~(\ref{ITF}) by adding three-body interactions. The method can be generalized to arbitrary interaction orders. This extension allows us to simulate problems such as quantum chemistry~\cite{LanyonQChem,Yung14,Poulin14,Babbush14}, as well as fermionic lattice models~\cite{Casanova12,Heras15}, by using the Jordan-Wigner mapping to map fermionic interactions into spin interactions. The Hamiltonian for $N$ spins including three-body interactions can be written as
\begin{equation}\label{CSC}
H = J\sum_{\langle jk \rangle}\sigma^z_j\sigma^z_k + G\sum_{\langle jkl \rangle}\sigma^z_j\sigma^z_k\sigma^z_l + B\sum_j\sigma^x_j,
\end{equation}
Here, we have added one collective interaction term with coupling constant $G$, which is the coupling among three nearest neighbour spins. This model can be simulated by enriching with additional gates the protocol for the Ising model in section~\ref{sec:3.1}. That is, together with single-qubit rotations along the $x$ direction, $X_j(\phi)=\exp(-i\phi\sigma^x_j)$, and two-qubit $zz$ gates, $ZZ_{jk}(\theta)=\exp(-i\theta\sigma^z_j\sigma^z_k)$, we also consider the combination of collective gates shown in Fig.~\ref{MSProtocol}. This will allow us to couple three qubits, $ZZZ_{jkl}(\beta)=\exp(-i\beta\sigma^z_j\sigma^z_k\sigma^z_k)$.

The collective spin interaction of this model can be decomposed into $(N-1)$ two-qubit gates and $2(N-2)$ three-qubit gates, and the transverse field is mimicked by $N$ single qubit rotations. Moreover, we notice that the digital error of the Trotter expansion in Eq.~(\ref{EqTrotter}) is reduced due to the fact that the interaction summands of the Hamiltonian commute with each other. The Trotter expansion for this model reads
\begin{equation}
e^{-iHt} \simeq \left(e^{-i\ t/s\ J\sum_{\langle jk \rangle}\sigma^z_j\sigma^z_k} e^{-i\ t/s\ G\sum_{\langle jkl \rangle}\sigma^z_j\sigma^z_k\sigma^z_l} e^{-i\ t/s\ B\sum_j\sigma^x_j}\right)^s,
\end{equation}
where
\begin{eqnarray}
& & \exp(-i\theta \sum_{\langle jk \rangle}\sigma^z_j\sigma^z_k) = e^{-i\theta\sigma^z_1\sigma^z_2}\ e^{-i\theta\sigma^z_2\sigma^z_3}\ \cdots\ e^{-i\theta\sigma^z_{N-1}\sigma^z_N}, \nonumber \\
& & \exp(-i\beta \sum_{\langle jkl \rangle}\sigma^z_j\sigma^z_k\sigma^z_l) = e^{-i\beta\sigma^z_1\sigma^z_2\sigma^z_3}\ e^{-i\beta\sigma^z_2\sigma^z_3\sigma^z_4}\ \cdots\ e^{-i\beta\sigma^z_{N-2}\sigma^z_{N-1}\sigma^z_N}, \nonumber \\
& & \exp(-i\phi\sum_j\sigma^x_j) = e^{-i\phi\sigma^x_1}\ e^{-i\phi\sigma^x_2}\ \cdots\ e^{-i\phi\sigma^x_N},
\end{eqnarray}
with $\theta=-Jt$, $\beta=Gt$ and $\phi= Bt$, $t$ being the simulated execution time.
The collective gate for three qubits can be decomposed into two-qubit gates, as in Fig.~\ref{MSProtocol}. Recently, the implementation of collective gates with a quantum bus has been proposed in superconducting circuits~\cite{MezzaMolmer}.

\begin{figure}
\sidecaption[t]
\includegraphics[scale=.67]{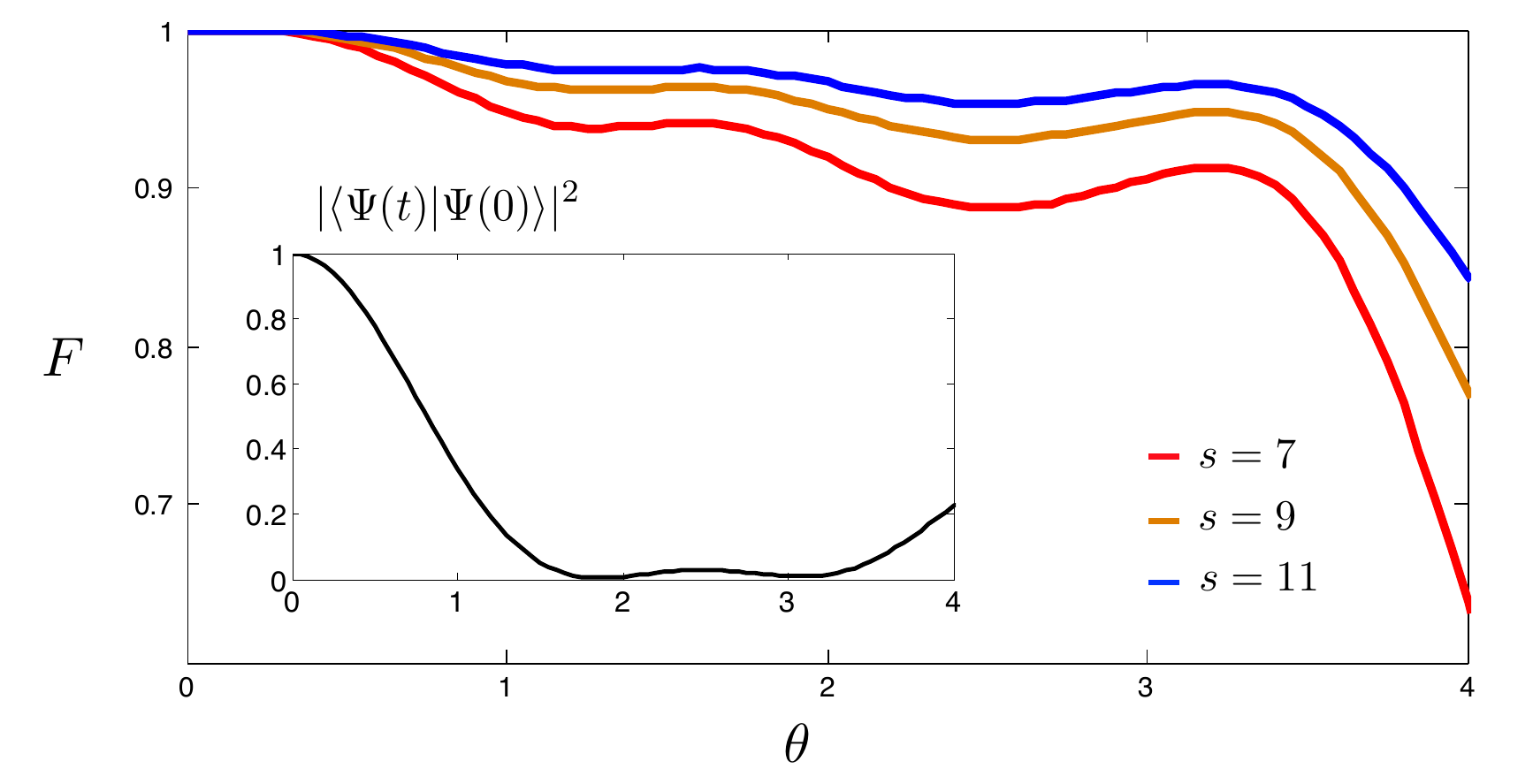}
%
%
\caption{Digital simulation of the extended Ising model with a transverse field and three-body terms for four spins in superconducting circuits, where $J=G=2\pi\times400$~MHz and $B=2\pi\times200$~MHz, for a phase of $\theta\equiv -Jt=4$. The plot shows the fidelity of the digitally evolved state with the ideally evolved one for different Trotter steps, $s=7,9,11$. The inset shows the overlap between the ideally evolved state with the initial state, that is, all qubits in $|0\rangle_z$ state.}
\label{CollectivePlot}       
\end{figure}

 In Fig.~\ref{CollectivePlot}, we plot a numerical simulation of the extended Ising model with higher-order terms and transverse field for several Trotter steps. The figure shows as in the previous examples how the simulated dynamics with digital methods becomes more accurate when compared with the exact one when the number of Trotter steps is increased.

\section{Conclusions}
In this article, we have proposed a digital quantum simulation of spin chains coupled to bosonic modes by means of circuit quantum electrodynamics architectures. We have presented a method for decomposing spin interactions and implementing them stroboscopically with available single and two-qubit gates. Furthermore, we have considered both circuit QED setups implementing capacitive couplings between superconducting qubits and transmission line resonators acting as quantum buses. We have exemplified our method with the quantum simulation of the Ising model with transverse field, a spin chain coupled to a bosonic field mode, and a many-body spin model with three-body terms, which are realized through a bosonic quantum bus. These results show that spin chains and bosonic field modes can be implemented efficiently with superconducting qubits. 

We thank Rami Barends for useful discussions and acknowledge support from Spanish MINECO FIS2012-36673-C03-02; Ram\'on y Cajal Grant RYC-2012-11391; UPV/EHU Project EHUA14/04, and two UPV/EHU PhD grants, Basque Government IT472-10; PROMISCE, and SCALEQIT EU projects.
\label{sec:4}

\end{document}